\documentclass[12pt]{article}

\usepackage{newtxtext,newtxmath}
\usepackage{graphicx}

\usepackage[letterpaper,margin=1in]{geometry}

\renewenvironment{abstract}
	{\quotation}
	{\endquotation}

\date{}

\makeatletter
\renewcommand{\fnum@figure}{\textbf{Figure \thefigure}}
\renewcommand{\fnum@table}{\textbf{Table \thetable}}
\makeatother

\usepackage{scicite}

\usepackage{url}

\def\scititle{
	Evidence for Quark Confinement in the Proton
}
\title{\bfseries \boldmath \scititle}

\author{
	Xiangdong Ji$^{1,2\ast}$,
	Gerald A. Miller$^{3\dagger}$,
    Chen Yang$^{2\ddagger}$\and
	\small$^{1}$Tsung-Dao Lee Institute and School of Physics and Astronomy, \\ \small Shanghai Jiao Tong University, Shanghai 201210, China.\and
	\small$^{2}$Department of Physics, University of Maryland, College Park, MD 20742, U.S.A.\and
    \small$^{3}$Department of Physics, University of Washington, Seattle, WA 98195-1560, U.S.A.\and
	\small$^\ast$Corresponding author. Email: xji@umd.edu%\and
}
\begin{document}

% Insert the title and author list
\maketitle

% Abstract, in bold
% There are strict length limits, and not all formats have abstracts.
% Consult the journal instructions to authors for details.
% Do not cite any references in the abstract.
\begin{abstract} \bfseries \boldmath
    The strong interaction is the fundamental force that holds quarks and the gluon force carriers together to form protons and neutrons and also binds the atomic nucleus. The theory governing quark-gluon interactions is Quantum Chromodynamics (QCD). A wide variety of experimental data  teaches us that quarks and gluons cannot be observed in isolation, a phenomenon known as confinement that is unique to QCD. But no one has used QCD to mathematically prove confinement. Here we show how to define and measure the force on quarks in the proton using available experimental data. Direct evidence for confinement is obtained because the force is found to be attractive and constant for a wide range of quark positions. This work guides future experimental efforts at  future Electron-Ion Colliders aimed at obtaining a rigorous quantitative understanding of confinement and the origin of nuclear mass.
\end{abstract}

% \section{Main}\label{sec1}
% The first paragraph of any Science paper does NOT have a heading
% Nor is it indented
\noindent
Rapid progress in understanding the role of the quark and gluon energy-momentum tensor (EMT) and the associated momentum conservation laws~\cite{Ji:1996ek,Ji:2025gsq,Tanaka:2018wea,Polyakov:2018exb}, along with the newly found ability to measure its properties~\cite{Burkert:2018bqq,Kumericki:2019ddg,Duran:2022xag,GlueX:2023pev}, has recently occurred. But the link between these properties and the underlying mystery of quark confinement has not been established. 

There are several important reasons for this. The difficulties can be understood by comparing the present situation with the history of the most prominent forces in our daily lives---gravity and electromagnetism. The inverse square laws of gravity and electricity were both discovered in table-top laboratory experiments by Cavendish and Coulomb using a torsion balance~\cite{coulomb1785premier,cavendish1879electrical}, but it is impossible to place quarks on a torsion balance and directly read the forces between them. This is because of the effects of confinement~\cite{Wilson:1974sk,Polyakov:1976fu,Greensite:2003bk} that prevent the observation of a free quark. Furthermore, in the small-sized world of the proton, the uncertainty principle precludes finding quarks at a definite position. To make matters even more difficult, the quarks within the proton move at speeds very close to those of light. Einstein's theory of relativity must be respected. % However, obtaining a direct measurement of the strong-interaction forces on quarks faces significant challenges because 

%discovery of 
  
%Unlike gravity and electromagnetism, the strong forces acting 
 
Thus, the fundamental strong forces acting between colored quarks have never been accessed by direct experimental measurement, but instead are postulated within a theory known as quantum chromodynamics (QCD) that operates in the sub-atomic quantum domain. The interactions between quarks separated by small distances, namely asymptotic freedom~\cite{Gross:1973id,Politzer:1973fx}, have been studied using high-energy scattering, but at larger distances our knowledge is largely limited to models of hadron spectroscopy~\cite{Eichten:1978tg,Godfrey:1985xj,QuarkoniumWorkingGroup:2004kpm} and models of static potentials for massive quarks~\cite{Bali:2000gf}. Finding a way to measure the forces on quarks that is directly connected to experimental measurements is therefore necessary to study inter-quark forces at large separations without using models. If this is accomplished, a test of the fundamental theory and understanding of the QCD confinement mechanism would be obtained.

In the following, we show how to overcome the various roadblocks mentioned above. The first step is to understand the proper definition of forces. This is done using the QCD EMT, the strong interaction version of the well-known Maxwell stress tensor used to understand electromagnetic forces. The nucleon matrix element of the EMT is defined in terms of kinematic factors and relativistically invariant form factors with the specific forms chosen so that the conservation of energy and momentum is respected~\cite{Ji:1996ek}. Then one needs to obtain the form factors from various experimental measurements~\cite{Ji:1996nm,Muller:1994ses,Radyushkin:1996ru} and fundamental numerical calculations known as lattice QCD~\cite{Wilson:1974sk}. This information allows a force density to be obtained. The force itself is obtained by dividing its density by the density of quarks~\cite{Hofstadter:1955ae}, with the latter also obtained from experimental measurements. All this is not yet enough. It is necessary to obtain an unambiguous way to define the position of a quark that respects the theory of relativity and the uncertainty principle. This is done by making evaluations in an infinite momentum frame~\cite{Dirac:1949cp,Weinberg:1966jm,Kogut:1969xa}. The net result is a well-defined expression for the force on quarks in a proton that can be evaluated using the relevant input information available.

\section{Force Density Form Factors in the Proton} \label{sec2}

We shall extend a method to extract the force on quarks that has been well-tested in non-relativistic, quantum mechanical systems~\cite{Ji:2025qax}. Force densities are obtained from the divergence of the EMT~\cite{landau1986theory}, which receives contributions from quarks, gluons, and an anomalous term~\cite{Ji:2025gsq}. For a stable system, the divergence of the total EMT vanishes, but the individual terms have non-vanishing divergences, giving rise to force components. Our interest here is in the forces on the quarks.
%In relativistic quantum field theories, particularly the nucleon in QCD, extracting the quark confinement force involves the quark force density and probability density in hadrons. These physical quantities can be expressed in terms of several form factors in hadrons in the following method. 

%On one hand, the quark force density is obtained by taking the divergence of the QCD quark EMT, including momentum densities and flows. 
% The conserved EMT is a consequence of spacetime translational symmetry, 
% \begin{equation}
%     \partial_\mu T^{\mu\nu} = 0
% \end{equation}
% In the QCD nucleon, this total EMT receives contributions from the kinetic motion of quarks $T^{\mu\nu}_q$ as well as the gluon tensor $T^{\mu\nu}_g$ and trace anomaly $T^{\mu\nu}_a$~\cite{Ji:2025qax}. 
% It has been shown in Ref.~\cite{Ji:2025qax} that the force density experienced by the matter particles is defined as 
% \begin{equation}
%     {\cal F}^j\equiv\partial_\mu T_K^{\mu j} = -\partial_\mu T_V^{\mu j}
% \end{equation}
% where $T_K^{\mu j}$ and $T_V^{\mu j}$ are the kinetic and interaction components. 
% In static systems, the momentum density is time-independent and thus the force density reduces to the divergence of the MCD. 
% In the QCD nucleon, the kinetic part is given by the quark EMT $T^{\mu\nu}_q$, while the interaction part contains both the gluon tensor and the trace anomaly contributions $T^{\mu\nu}_V = T^{\mu\nu}_g + T^{\mu\nu}_a$. 
%divergence of the quark EMT at the operator level yields, 
The density of the forces acting on the quarks ($q$) is given by,
\begin{equation}
    {\cal F}_q^i \equiv \partial_\mu T^{\mu i}_q = g\bar{\psi}\gamma_{\mu}F^{\mu i}\psi = \sum_{c=1}^8g(\rho_c \vec{E}_c + \vec{j}_c\times\vec{B}_c)^i  \ , 
\label{fdens}\end{equation}
where $T_q^{\mu\nu}$ is the EMT related to the quarks, $c$ is the color index, $i$ represents a spatial variable, $g$ is the strong coupling constant, $\{\rho_c,\vec{j}_c\}$ are the quark color-charge density and current and $\{\vec{E}_c,\vec{B}_c\}$ are the chromo-electromagnetic fields. The QCD equation of motion allows us to show that this quantity is equal to the gluon version of the Lorentz force law, or the color-Lorentz force~\cite{Tanaka:2018wea,Polyakov:2018exb}, that is familiar from electromagnetism. 
%. This definition acquires a clear physical interpretation as the color-Lorentz force density acting on quarks. 
% Additionally, the above definition shows that the force density is only related to the non-conserved (non-zero divergence) part in the EMT. 

Nucleon momentum densities and flows are measured through high-energy experiments as the matrix elements of the EMT between eigenstates of momentum, parametrized in terms of kinematic terms and EMT form factors, expressed as $\{A,B,C,\bar{C}\}$~\cite{Ji:1996ek}. These are experimentally accessible through exclusive processes such as deeply virtual Compton scattering (DVCS) and deep exclusive meson production (DVMP)~\cite{Ji:1996ek,Ji:1996nm,Muller:1994ses,Radyushkin:1996ru,Collins:1996fb,Diehl:2003ny,Belitsky:2005qn,Burkert:2018bqq,Kumericki:2019ddg,Duran:2022xag,GlueX:2023pev,Guo:2025jiz,Guo:2025muf}. The EMT form factors can also be computed using first-principles lattice-QCD (LQCD) calculations~\cite{Hackett:2023rif}. 
%~\cite{Hagler:2003jd,Gockeler:2003jfa,LHPC:2007blg,Hackett:2023rif}. 

%only related to the non-conserved (nc) part of the quark EMT and associated form factor g
The  force density, $ {\cal F}_q^j$, is obtained from the term in the quark part of the EMT with a non-vanishing divergence; these are denoted as non-conserved (nc). The terms with vanishing divergence give no contribution to the force density. This density is obtained from~\cite{Ji:1996ek}, 
\begin{align}
    \langle P^{\prime}|T_{q}^{\mu\nu}|P\rangle_{\rm nc} =\bar{U}(P^{\prime}) \left[-\frac{1}{4}g^{\mu\nu}MG_{s,q}(q^2)\right]U(P) \ , \label{eq:ncFF1}
\end{align}
where $U$ and $\bar U$ are proton momentum-spin wave functions, $M$ is the proton mass, $q^{\mu}=P^{\prime\mu}-P^{\mu}$ is the momentum transfer and $G_{s,q}(q^2)$ is the quark scalar form factor~\cite{Ji:2021mtz},
\begin{align}
    G_{s,q}(q^2)=A_{q}(q^2)+B_{q}(q^2)\frac{q^{2}}{4M^{2}}-C_{q}(q^2)\frac{3q^{2}}{M^{2}} \ ,\label{eq:gs}
\end{align} 
with $q^2=(P'-P)^2$. The form factor $B_q(q^2)$ is relatively small and is therefore neglected in the following calculations.
%Thus, obtaining the quark force density only requires this quark scalar form factor. Notably, 

\section{Relativistic Position Variable Respecting Uncertainty Principle}

Experimental data and LQCD only measure the matrix elements of the EMT and force density ${\cal F}_q^j$ between momentum states, as shown in Eq.~(\ref{eq:ncFF1}), but such states do not have a well-defined position. Furthermore, the different momenta of the initial and final states lead to different internal dynamics and structures due to relativity, while a true density is defined by a matrix element between the same states. 

To handle this, the proton state vector, for both the initial and final states, is taken to be an infinite superposition of states with fixed infinite momentum in one direction ($z$) and any momentum transverse to $z$: we use an infinite momentum frame~\cite{Dirac:1949cp,Weinberg:1966jm,Kogut:1969xa}. In this frame, the transverse momentum factorizes out of the Hamiltonian such that states of any momentum transverse to $z$ have the same internal dynamics. This superposition yields a state of well-defined transverse position that can be taken to be the origin, so that transverse positions relative to the origin are defined~\cite{Burkardt:2002hr,Miller:2007uy,Miller:2025zte}. Details are provided in Supplementary Materials.

In the 4-dimensional divergence appearing in Eq.~(\ref{fdens}), two of these derivatives vanish when matrix elements are taken between the proton state vector. This is because the energy transfer vanishes and the longitudinal momentum is  fixed at an infinite value. Thus, we are left with only the transverse parts of the divergence. This means that the force density is simply
\begin{align}
    \vec{\cal F}_{q}(\vec{r}_\perp) = (M/4)\vec{\nabla}_{\perp}G_{s,q}(\vec{r}_{\perp}) + {\cal O}\left(\frac{1}{(P^+)^3}\right)\ .
\label{good}\end{align}
The correction terms of order of the inverse cube of the infinite longitudinal momentum are safely ignored. See Eqs.~(\ref{S11}) and (\ref{S12}) in Supplementary Materials.

% To generalize this approach to relativistic hadron systems, we then analyze the IMF and justify the physical interpretation of spatial densities. 
% and naturally decomposes into an attractive contribution from the QCD trace anomaly and a repulsive contribution from the gluon tensor. 

% Research Articles and Reviews split the text into sections using headings
% Use a short (up 6 words) descriptive phrase, not generic 'Results' or 'Conclusions'
% Most other formats do not have headings, see the journal instructions to authors for details

\section{Measuring the Strong Forces acting on Quarks in the Proton}

The force on the quarks is the force density, Eq.~(\ref{good}), divided by the quark density, $\rho_q(r_\perp)$. The function $\rho_q(r_\perp)$ can be obtained, using the infinite momentum frame discussed above, by taking the two-dimensional Fourier transform of the sum of the proton and neutron Dirac form factors $3[F_1^p(q^2)+F_1^n(q^2)]$~\cite{Miller:2007uy}, where the factor $3$ reflects the number of valence quarks. Therefore, the color-Lorentz force on quarks is given by
\begin{equation}
    \vec{F}_{q}(\vec{r}_\perp) \equiv \frac{\vec{\cal F}_{q}(\vec{r}_\perp)}{\rho_q(\vec{r}_\perp)} = \frac{(M/4)\vec{\nabla}_{\perp} G_{s,q}(\vec{r}_{\perp})}{3[F_1^p(\vec{r}_\perp)+F_1^n(\vec{r}_\perp)]} \ ,
    \label{big}
\end{equation}
where the spatial densities are obtained by Fourier-Bessel transformations of the various form factors. See Supplementary Materials for details. Due to rotational symmetry in the transverse plane, the force lies exclusively along the radial direction. 
%Methods.~\ref{secA2} and~\ref{secA3} for details. 

Eq.~(\ref{big}) shows how to precisely determine the force once all the form factors are known for all a large range of values of the momentum transfer. The accuracy of our force determination depends on the current incomplete nature of this knowledge. 

\begin{figure}[htb]
    \centering
    \includegraphics[width=0.65\linewidth]{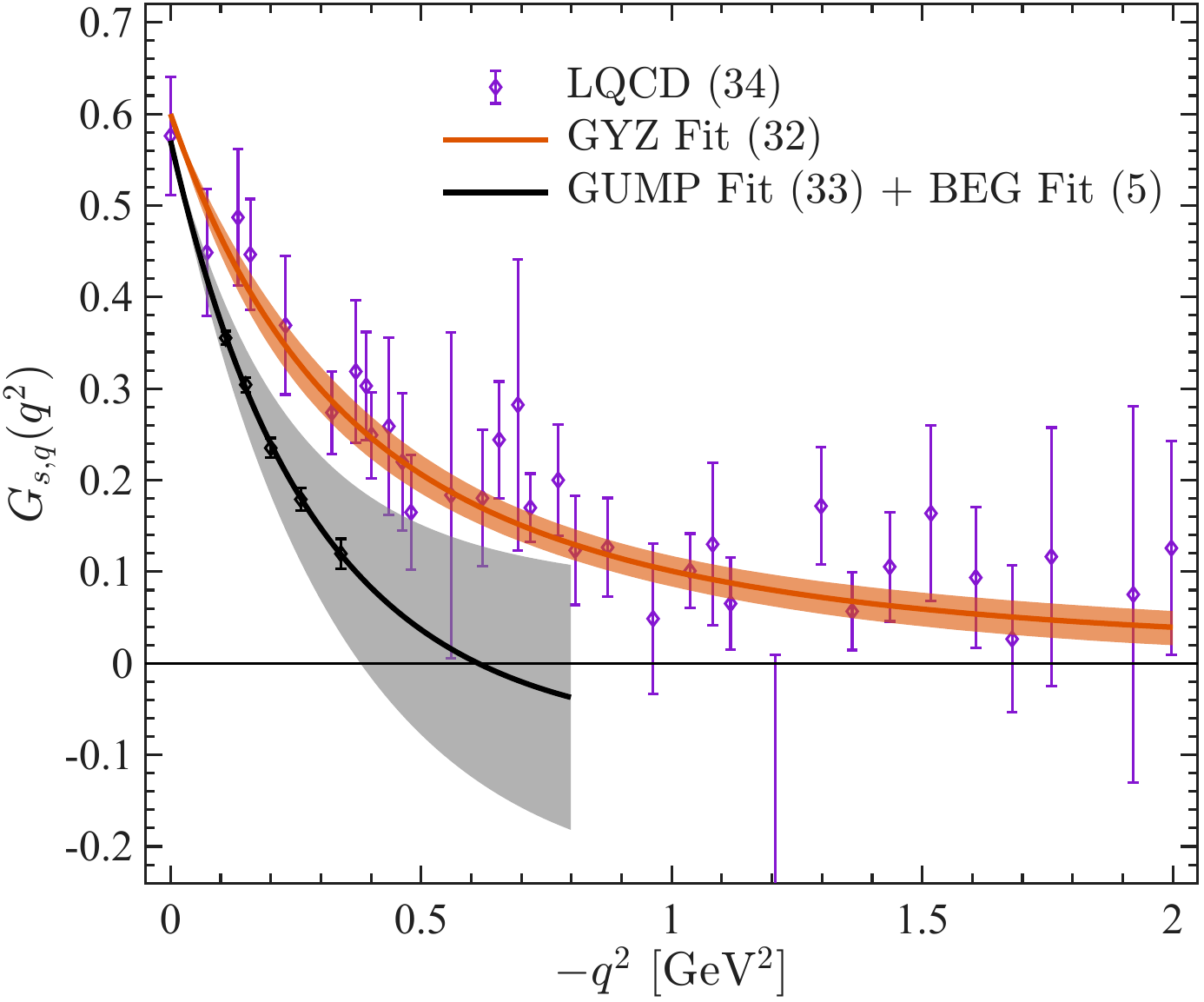}
    \caption{
    \textbf{The quark scalar form factor $G_{s,q}(q^2)$ obtained from LQCD calculations and experimental data.} Direct LQCD calculations are shown as purple error bars~\cite{Hackett:2023rif}. The GYZ fit (orange curve and band) represents the Bayesian inference of LQCD (main) and $J/\psi$ photoproduction data~\cite{Guo:2025jiz}. The black curve and band exhibits the experimental information from GUMP fit and BEG's dispersive analysis of DVCS data~\cite{Burkert:2018bqq,Guo:2025muf}.
    }
    \label{fig:Gs}
\end{figure}

We perform a comprehensive assessment of the state-of-the-art global phenomenological analyses, including both LQCD and experimental inputs to extract the quark scalar form factor~\cite{Burkert:2018bqq,Kumericki:2019ddg,Hackett:2023rif,Guo:2025jiz,Guo:2025muf}; the results are summarized in Fig.~\ref{fig:Gs}. The LQCD calculations of the EMT form factors, (using pion mass $m_\pi = 169$~MeV, and lattice spacing $a=0.091$~fm), are shown as data points with purple error bars~\cite{Hackett:2023rif}. The $A_q(q^2)$-form factor, which appears as part of $G_{s,q}(q^2)$, is rescaled so that $A_q(0)$ of LQCD is consistent with the Coordinated Theoretical-Experimental Project on QCD (CTEQ)  extraction~\cite{Hou:2019efy}. The GYZ fit (orange curve with its uncertainty band) is a Bayesian analysis driven primarily by lattice constraints, with only weak sensitivity to near-Threshold $J/\psi$ photo-production measurements~\cite{Guo:2025jiz}. The experimental information is mainly reflected by the black curve and band, obtained by combining the global extraction of generalized parton distributions (GUMP fit)~\cite{Guo:2025muf} and the dispersive analysis of DVCS data (BEG fit)~\cite{Burkert:2018bqq}. Due to the limited kinematic range available for current experiments, the inferred scalar form factor is poorly constrained at large values of momentum transfer, $|q^2|$. See Supplementary Materials section for numerical details and error analyses. 
%, except that $A_q(0)$ is fixed by experimental input from CTEQ extraction. 

\begin{figure}[htb]
    \centering
    \includegraphics[width=0.65\linewidth]{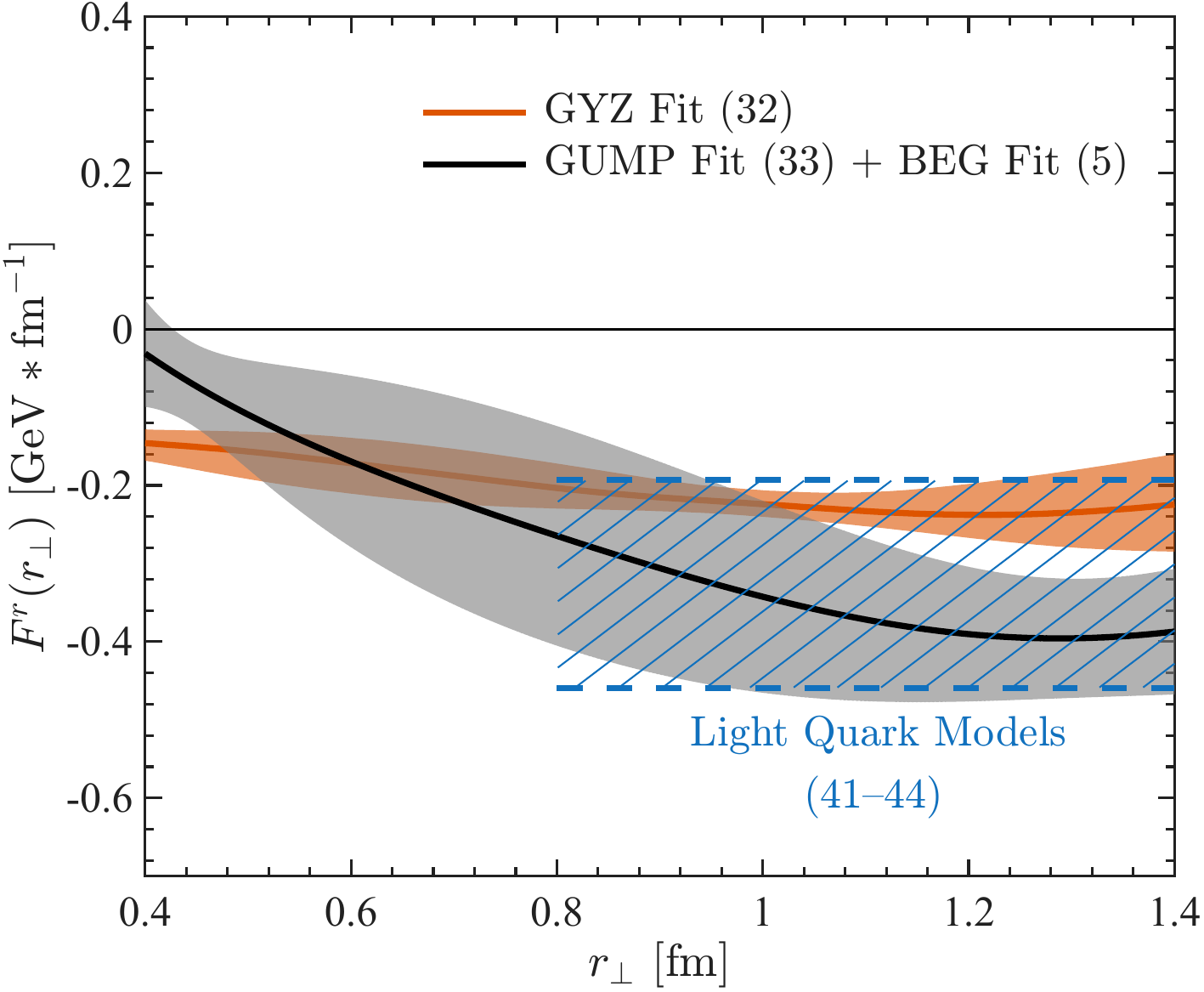}
    \caption{
    \textbf{Strong force on quarks in the proton in the transverse plane of the IMF.} The forces are obtained using the latest global analyses of the quark momentum current form factors~\cite{Burkert:2018bqq,Guo:2025jiz,Guo:2025muf}. The figure shows the total force from experimental data (black curve and band) and from the fit dominated by LQCD calculations (orange curve and band). The blue hatched band shows the string tensions from several relativistic quark models~\cite{LealFerreira:1979xq,Eich:1983kg,Dziembowski:1996cv,Glozman:1997ag}.
    }
    \label{fig:ForceIMF}
\end{figure}

The force acting on the proton's quarks in the IMF, using Eq.~(\ref{big}) is displayed in Fig.~\ref{fig:ForceIMF}. The denominator of Eq.~(\ref{big}) is obtained from the Dirac proton and neutron form factors of Ref.~\cite{Kelly:2004hm}. The forces extracted from the GYZ fit and GUMP \& BEG fits, together with the $1\sigma$ deviations, are shown as the orange and black curves and associated shaded areas. For comparison, we also show the constant confining force inferred from several relativistic quark models as the blue hatched band~\cite{LealFerreira:1979xq,Eich:1983kg,Dziembowski:1996cv,Glozman:1997ag}. 

The experimental results from GUMP \& BEG fits provide strong evidence for a net confining force acting on quarks. At intermediate distances, specifically $r_\perp=1.0\sim1.4$~fm, where the uncertainties are relatively small, the total force acting on the quarks extracted from the experimental data is consistent with the QCD linear potential. Averaging over this interval yields $\bar{F}_q = -0.382(92)$~GeV/fm, where the uncertainty corresponds to a 1$\sigma$ deviation. On the other hand, the GYZ fit, dominated by the LQCD data, produces a mean value $\bar{F}_q = -0.217^{+0.016}_{-0.015}$~GeV/fm with uncertainties $1\sigma$, for the range  $r_{\perp}=0.7\sim1.2$~fm. The GYZ fit exhibits relatively small uncertainties, obtained from a Monte Carlo replica method. These values are also compatible with relativistic quark models. The negative value of the force denotes an attractive potential on the quark because the force is the negative gradient of the potential.

Extending the curves to larger values of $r_\perp$ than shown in Fig.~\ref{fig:ForceIMF} would involve a significant increase in the uncertainties, particularly for experimental extraction. This illustrates the challenges in measuring the EMT form factors. These  can only be accessed through processes sensitive to generalized parton distributions~\cite{Ji:1996ek,Muller:1994ses}. Measuring these processes generally  requires data from high-energy and high-luminosity facilities with detectors capable of precise and complete mapping of scattering events. Therefore, data from the future Electron-Ion colliders~\cite{AbdulKhalek:2021gbh,Anderle:2021wcy} are crucially needed to improve the precision of EMT form factors and thereby achieve a rigorous quantitative understanding of color confinement. 

\section{Discussion and Summary} 

The approximately constant strong force on the component quarks observed far away from the proton center that we find is consistent with the known QCD linear potential acting between heavy quarks. This confinement force is about half the QCD string tension $\sigma\approx 0.9\ {\rm GeV/fm}$ modeled from heavy quarkonium. This reduction is expected because the motion of the light quarks that make up the proton spreads out the postulated flux tube between the quarks~\cite{Bali:2000gf} as is also indicated by the hopping expansion~\cite{Rothe:1992nt}. A lower string tension for light quarks is also consistent with the findings in light mesons~\cite{Kawanai:2011xb}. On the other hand, in the asymptotically large-distance regime, the force on quarks shall decay exponentially to zero. From the perspective of dispersive analysis, this behavior follows from the fact that the lightest pole governing the scalar form factors in the positive $q^2$ region, corresponding to a $0^{++}$ glue-ball state~\cite{Mamo:2022eui}, has a mass scale larger than that controlling the Dirac form factors, which is given by the $\rho$ meson mass~\cite{Mergell:1995bf}. In a more intuitive picture, the exponential decay reflects flux-tube breaking for a widely separated $q\bar{q}$ pair---consistent with the well-known plateau of the static potential at large separations observed in LQCD calculations~\cite{Bali:2000gf}. 

We note that the GYZ fit, based mainly on LQCD results, yields a weaker confining force than the one inferred from the experimental data. This difference is plausibly attributable to various lattice artifacts such as excited state contamination, the large  pion mass, and the finite lattice spacings used, as $A_q(0)$ shows a significant deviation from the data. 
%Similar deviations have also been  observed in comparing the electromagnetic form factors between LQCD and experiments~\cite{Alexandrou:2018sjm}. 

In summary, we show how to directly determine the strong force on quarks within hadrons using momentum current distributions and associated form factors in Eq.~(\ref{big}) accessible through high-energy experiments and lattice QCD calculations. By applying our approach to the proton in the infinite-momentum frame, we obtain the strong-interaction force acting on quarks and strong evidence for quark confinement. The total confinement force is consistent with the linear potential nature of QCD, although with significant experimental uncertainties. Obtaining future data from the electron-ion colliders and other experiments will be crucial for reducing uncertainties at small and large distances and finally obtaining a precise tomography of the quark confinement forces. 
%Several quantum mechanical systems provide intuitive illustrations and guides to extracting the interacting forces. 
% {\color{blue} Flavor decomposed forces.}

%%%%%%%%%%%%%%%% REFERENCES %%%%%%%%%%%%%%%

\clearpage % Clear all remaining figures and tables then start a new page

% The list of references goes after the main text and before the acknowledgements
% When preparing an initial submission, we recommend you use BibTeX, like this:
%
\bibliographystyle{sciencemag}
\bibliography{ref}

%%%%%%%%%%%%%%%% ACKNOWLEDGEMENTS %%%%%%%%%%%%%%%

\section*{Acknowledgments}

The authors thank Z.-E. Meziani for continually  asking questions about possible experimental signatures for color confinement, Yushan Su and Yijie Wang for useful discussions, and Yuxun Guo for providing the data for the parametrization of EMT FFs. 
\paragraph*{Funding:}
X. Ji and C. Yang are partially supported by the Maryland Center for Fundamental Physics (MCFP). G.A. Miller is partially funded by the U. S. Department of Energy, Office of Science, Office of Nuclear Physics under Grant No. DE-SC0026252.

\paragraph*{Author contributions:}
X. Ji and G.A. Miller conceived the project and supervised the research. C. Yang carried out the theoretical derivations and numerical calculations. All authors discussed the physics results, interpreted the data and contributed to writing the manuscript.

\paragraph*{Competing interests:}
The authors declare no competing interests.

\paragraph*{Data and materials availability:}
Correspondence and requests for data and materials should be addressed to X.J., G.A.M., or C.Y.

%%%%%%%%%%%%%%%% SUPPLEMENT LIST %%%%%%%%%%%%%%%

% List the contents of your Supplementary Materials, including the numbers of any
% supplementary figures, tables, external data files etc. and any references that are
% cited only in the supplement. In this example, refs. 7-8 are cited only in the supplement.
% Fill out your numbers accordingly and delete any lines that aren't applicable.
\subsection*{Supplementary materials}
Materials and Methods\\

%%%%%%%%%%%%%%%% END OF MAIN TEXT %%%%%%%%%%%%%%%

\newpage

%%%%%%%%%%%%%%%% START OF SUPPLEMENT %%%%%%%%%%%%%%%

% Figures, tables, equations and pages in the supplement are numbered S1, S2 etc.
\renewcommand{\thefigure}{S\arabic{figure}}
\renewcommand{\thetable}{S\arabic{table}}
\renewcommand{\theequation}{S\arabic{equation}}
\renewcommand{\thepage}{S\arabic{page}}
\setcounter{figure}{0}
\setcounter{table}{0}
\setcounter{equation}{0}
\setcounter{page}{1} % not 0 as \newpage already started a supplementary page
% References continue the numbering from the main text.

% \appendix
% \section*{Methods}
\begin{center}
\section*{Supplementary Materials for\\ \scititle}

% Author list for the supplement
% Indicate the corresponding authors, but do NOT include institutions here
% It would be nice if the template auto-generated this, but doing so is complicated...
Xiangdong Ji$^{\ast}$,
Gerald A. Miller,
Chen Yang\\
\small$^\ast$Corresponding author. Email: xji@umd.edu

\end{center}

\subsection*{Force Density Form Factors in the Proton} \label{secA2}

In relativistic quantum field theories, particularly the nucleon in QCD, extracting the quark confinement force involves the quark force density and probability density in hadrons. These physical quantities can be expressed in terms of several form factors in hadrons using the following method. The key expression is Eq.~(\ref{big}).

The numerator of Eq.~(\ref{big}), the quark force density, is obtained by taking the divergence of the QCD quark EMT, including momentum densities and flows. 
% The conserved EMT is a consequence of spacetime translational symmetry, 
% \begin{equation}
%     \partial_\mu T^{\mu\nu} = 0
% \end{equation}
% In the QCD nucleon, this total EMT receives contributions from the kinetic motion of quarks $T^{\mu\nu}_q$ as well as the gluon tensor $T^{\mu\nu}_g$ and trace anomaly $T^{\mu\nu}_a$~\cite{Ji:2025qax}. 
% It has been shown in Ref.~\cite{Ji:2025qax} that the force density experienced by the matter particles is defined as 
% \begin{equation}
%     {\cal F}^j\equiv\partial_\mu T_K^{\mu j} = -\partial_\mu T_V^{\mu j}
% \end{equation}
% where $T_K^{\mu j}$ and $T_V^{\mu j}$ are the kinetic and interaction components. 
% In static systems, the momentum density is time-independent and thus the force density reduces to the divergence of the MCD. 
% In the QCD nucleon, the kinetic part is given by the quark EMT $T^{\mu\nu}_q$, while the interaction part contains both the gluon tensor and the trace anomaly contributions $T^{\mu\nu}_V = T^{\mu\nu}_g + T^{\mu\nu}_a$. 
The divergence of the quark EMT at the operator level yields 
\begin{equation}
    {\cal F}_q^j
    \equiv \partial_\mu T^{\mu j}_q = g\bar{\psi}\gamma_{\mu}F^{\mu j}\psi = \sum_{c=1}^8g(\rho_c \vec{E}_c + \vec{j}_c\times\vec{B}_c)^j  \ ,
\end{equation}
where we have used the QCD equation of motion. The final expression is the color-Lorentz force density acting on quarks.
% Additionally, the above definition shows that the force density is only related to the non-conserved (non-zero divergence) part in the EMT. 

The nucleon momentum densities and flows are measured through high-energy experiments as the matrix elements of the QCD EMT in the momentum states, parametrized in terms of several EMT form factors. The above definition indicates that the force density is only related to the non-conserved (nc) part (the part with non-zero divergence) of the quark EMT and the associated form factor given by~\cite{Ji:1996ek}, 
\begin{align}
    \langle P^{\prime}|T_{q}^{\mu\nu}|P\rangle_{\rm nc} =\bar{U}(P^{\prime}) \left[-\frac{1}{4}g^{\mu\nu}MG_{s,q}(q^2)\right]U(P) \ , \label{eq:ncFF}
\end{align}
where $M$ is the proton mass, $q^{\mu}=P^{\prime\mu}-P^{\mu}$ is the momentum transfer, and $G_{s,q}(q^2)$ is the quark scalar form factor,
\begin{align}
    G_{s,q}(q^2)=A_{q}(q^2)+B_{q}(q^2)\frac{q^{2}}{4M^{2}}-C_{q}(q^2)\frac{3q^{2}}{M^{2}} \ . 
\end{align} 
Thus, obtaining the quark force density only requires this quark scalar form factor. The $B_q(q^2)$-form factor is relatively small and is thus neglected. 

The denominator  of Eq.~(\ref{big}), the quark probability density is approximated by using the Dirac proton and Dirac neutron  form factors in the following way. In the IMF, proton and neutron's charge densities are related to the Dirac form factors~\cite{Miller:2007uy}, which can be approximated using the $u$- and $d$-quark densities in the proton, 
\begin{equation}
    \rho_p = F_1^p \approx \frac{2}{3}\rho_u - \frac{1}{3}\rho_d, ~~~~ \rho_n = F_1^n \approx \frac{2}{3}\rho_d - \frac{1}{3}\rho_u \ ,
\end{equation}
where the second equation is obtained using isospin symmetry. Using these two equations, one can approximate the quark probability density in the proton as  
\begin{equation}
    \rho_q \approx \rho_d+\rho_u \approx 3(F_1^p+F_1^n) \ .
    \label{rhod}
\end{equation}
The quark density is extracted by taking the two-dimensional Fourier transform of these form factors as discussed in the following section. In summary, one uses  Eqs.~(\ref{eq:ncFF}) and (\ref{rhod}) as inputs to the numerator and denominator of Eq.~(\ref{big}).

\subsection*{Spatial Densities in the Infinite-Momentum Frame}\label{secA3}%

Since form factors are defined in the momentum space,   exploring the related spatial distributions requires care. In quantum field theories, spatial distributions are defined as the expectation values of local operators ${\cal O}(x)$ in a localized single-hadron state $|\Psi_\sigma\rangle$. By constructing the localized state using a superposition of hadron momentum eigenstates with a Gaussian wave packet, one can derive the relation between the form factors and the spatial distributions. 

The time dependence $x^+$ naturally vanishes in the infinite momentum frame defined by $P^+\to\infty$, aligning with Feynman's motivation for creating the parton model. Then a Gaussian wave packet with spatial width $\sigma$ in the transverse plane is used to define the localized state $|\Psi_\sigma\rangle$,
\begin{equation}
    \langle p^+,\vec{p}_\perp|\Psi_\sigma\rangle = \sqrt{2\pi}(2\sigma)e^{-\sigma^2\vec{p}_\perp^2}\sqrt{2p^+(2\pi)\delta(p^+-P^+)}.
\end{equation}
The inverse spatial width must be much smaller than the momentum of the hadron $1/\sigma \ll P^+$. The limit $\sigma\to0$ is to be taken at the end of the calculation.  The 2D spatial density is then obtained as the expectation value of the operator in this localized state, related to the form factors by~\cite{Miller:2025zte},
\begin{align}
    \rho_{\cal O}(\vec{r}_\perp) \equiv& \lim_{P^+\to\infty} \int{\rm d}r^{-}\langle \Psi_\sigma|\hat{{\cal O}}(r^{+},r^{-},\vec{r}_{\perp})|\Psi_\sigma\rangle \nonumber \\
    =&(2\pi)(2\sigma)^2\int\frac{{\rm d}^2 \vec{P}_{\perp}}{(2\pi)^2}e^{-2\sigma^2\vec{P}_{\perp}^2} \int\frac{{\rm d}^{2}\vec{q}_{\perp}}{(2\pi)^{2}}\frac{\langle P^+,\vec{p}_{\perp}^{\prime}|\hat{{\cal O}}(0)|P^+,\vec{p}_{\perp}\rangle}{2P^{+}}e^{-\sigma^{2}\vec{q}_{\perp}^{2}/2}e^{-i\vec{q}_{\perp}\cdot\vec{r}_{\perp}}  \ ,
\end{align}
where $\vec{P}_{\perp}$  is   the average transverse  momentum, $\vec{q}_{\perp}=\vec{p'}_{\perp}-\vec{p}_{\perp}$  is the momentum transfer in the transverse plane and the integral over $r^-$ ensures that the initial and final states have the same $P^+$. If the matrix elements only depend on the momentum transfer $\vec{q}_{\perp}$, one can integrate over  the  momenta and take the local limit $\sigma\to0$ in the above formula, and now the Fourier transformation of the matrix elements acquires  clear physical interpretations as the 2D spatial densities, 
\begin{equation}
    \lim_{\sigma\to0}\rho_{\cal O}(\vec{r}_\perp) = \frac{M}{P^+}\int\frac{{\rm d}^{2}\vec{q}_{\perp}}{(2\pi)^{2}}\frac{\langle P^+,\vec{p}_{\perp}^{\prime}|\hat{{\cal O}}(0)|P^+,\vec{p}_{\perp}\rangle}{2M}e^{-i\vec{q}_{\perp}\cdot\vec{r}_{\perp}}\label{dens}  \ ,
\end{equation}
This is the case for the proton and neutron's Dirac form factors $F^p_1(q^2)$ and $F^n_1(q^2)$. Using the above formula, the quark probability density on the transverse plane $3[F_1^p(\vec{r}_\perp)+F_1^n(\vec{r}_\perp)]$ is obtained by Fourier-Bessel transforming the Dirac form factors of the proton and neutron  $F_1^p(-\vec{q}^2_\perp)$ and $F_1^n(-\vec{q}^2_\perp)$.
% {\color{red} This part needs to be modified according to Chen's work showing the bad terms go away}

However, the EMT operators also contain terms in the matrix elements that depend on both $\vec{P}_{\perp}$ and $\vec{q}_{\perp}$, which, after integrating over the average transverse momentum, contribute as a ratio between the Compton wavelength $1/M$ and the localization scale $\sigma$. These terms are  divergent in the localized wave packet limit $\sigma\to0$, and therefore could render the interpretation as 2D spatial densities ambiguous. 

Fortunately, these physically ambiguous terms appear in the EMT operators as the conserved contributions, which, after taking the divergence to obtain the force density, precisely vanish. The non-conserved part is only related to the quark scalar form factor $G_{s,q}(q^2)$ contributing as a diagonal term proportional to $g^{\mu\nu}$. Thus, the 4-dimensional divergence reduces to the 2-dimensional divergence in the transverse plane and is independent of the average momentum $\vec{P}_\perp$. We omit the higher-twist factor $M/P^+$ in the transverse EMT, and the force density is thus obtained as,
\begin{align}
    {\cal F}^j_{q} (\vec{r}_\perp) &\equiv \partial_\mu \langle T_{q}^{\mu j} \rangle_{\rm nc} (\vec{r}_\perp) \nonumber \\
    &= \nabla^i_\perp \langle T_{q}^{ij} \rangle_{\rm nc} (\vec{r}_\perp) = \frac{M}{4}\nabla^j_{\perp}G_{s,q}(\vec{r}_\perp),
\end{align}
where $i,j=x,y$ and the subscript ${\rm nc}$ represents the non-conserved part. 

One may also explicitly examine this cancellation by calculating the divergence of the EMT,
\begin{equation}
    \partial_\mu \langle T_{q}^{\mu j} \rangle=\partial_- \langle T_{q}^{-j} \rangle+\partial_+ \langle T_{q}^{+ j} \rangle+\partial_i \langle T_{q}^{ij} \rangle \ . 
\end{equation}
Since the initial and final states have the same $P^+$, the $x^-$ dependence vanishes and thus the first term does not contribute. The other two components can also be computed directly, by taking the limit $P^+\to\infty$ in the final step, 
\begin{align}
    \partial_{+}\langle T_{q}^{+j}\rangle  &= \frac{1}{4\sigma^{2}P^{+}}\left[-\nabla_{\perp}^{j}\int\frac{{\rm d}^{2}\vec{q}_{\perp}}{\left(2\pi\right)^{2}}e^{-\sigma^{2}\vec{q}_{\perp}^{2}/2}e^{-i\vec{q}_{\perp}\cdot\vec{x}_{\perp}}A_{q}\left(-\vec{q}_{\perp}^{2}\right)\right]+{\cal O}\left(\frac{1}{\left(P^{+}\right)^{3}}\right) \ , \label{S11}\\
    \partial_{i}\langle T_{q}^{ij}\rangle  &= \frac{1}{4\sigma^{2}P^{+}}\left[+\nabla_{\perp}^{j}\int\frac{{\rm d}^{2}\vec{q}_{\perp}}{\left(2\pi\right)^{2}}e^{-\sigma^{2}\vec{q}_{\perp}^{2}/2}e^{-i\vec{q}_{\perp}\cdot\vec{x}_{\perp}}A_{q}\left(-\vec{q}_{\perp}^{2}\right)\right] + {\cal O}\left(\frac{1}{\left(P^{+}\right)^{3}}\right) + \partial_{i}\langle T_{q}^{ij}\rangle_{\rm nc} \ .
\label{S12}\end{align}
Remarkably, the sum of the two divergent terms in the $\sigma\to0$ limit  cancel to order $({1/P^+})^3$, and one is left with the 2D derivative over the scalar form factor, which is free of ambiguities. 

\subsection*{Numerical Demonstration}\label{secA4}

The numerical demonstration utilizes the parametrization of form factors, where the parameters are obtained from the latest global analysis including both experimental data and lattice QCD calculations. 

The proton and neutron's Dirac form factors are related to the electric and magnetic form factors as, 
\begin{equation}
    F_1^{(p,n)}(q^2) = \frac{G_E^{(p,n)}(q^2)+\tau G_M^{(p,n)}(q^2)}{1+\tau} \ ,
\end{equation}
where $\tau=-q^{2}/4M^{2}$. These electric and magnetic form factors are parametrized using the formula,
\begin{equation}
    {\mathbb G}(q^{2})=\frac{1+\sum_{k=1}^{n}a_{k}\tau^{k}}{1+\sum_{k=1}^{n+2}b_{k}\tau^{k}},~~~G_E^n(q^2)=\frac{A\tau}{1+B\tau}\frac{1}{(1-q^{2}/\Lambda^2)^{2}} \ ,
\end{equation}
where $n=1$, the form factors are ${\mathbb G}=\{G_E^p,G_M^p/\mu_p,G_M^n/\mu_n\}$, $\mu_p,\mu_n$ are the magnetic moments of the proton and neutron, the parameter is $\Lambda^2=0.71{\rm GeV}^{2}$ and the other parameters $\{a_k,b_k,A,B\}$ can be found in Ref.~\cite{Kelly:2004hm}. Moreover, uncertainties in the Dirac form factors are small and thus neglected. 

In the GYZ fits, the EMT FFs are parameterized using the multipole ansatz with two parameters $\{\mathbb{F}(0),M_{\mathbb{F}}\}$ as
\begin{equation}
    \mathbb{F}(q^2) = \mathbb{F}(0)( 1-q^2/M_{\mathbb{F}}^2 )^{-\alpha} \ ,
\end{equation}
where we choose $\alpha=2$ (dipole) for the $A_{q}(q^2)$ form factor and $\alpha=3$ (tripole) for the $C_{q}(q^2)$ form factor. The $B_q(q^2)$-form factor is relatively small and will be neglected. By performing the Fourier transformation with respect to the momentum transfer in the IMF $q^2=-\vec{q}^2_{\perp}$, one obtains the spatial distributions $\mathbb{F}(\vec{r}_\perp)$. The scalar form factor in Eq.~(\ref{eq:gs}) then becomes
\begin{equation}
    G_{s,q}(\vec{r}_\perp)=A_{q}(\vec{r}_\perp)-\frac{3}{M^{2}}\nabla^2_\perp C_{q}(\vec{r}_\perp)
\end{equation}
 We use $N=2\times 10^4$ parameter sets obtained from the Monte-Carlo Markov Chain (MCMC) sampling method~\cite{Guo:2025jiz} to estimate the uncertainties in the extracted scalar form factor and the force,. We then propagate these MCMC samples utilizing the form factor and force formulas in terms of these parameters, through which the full parameter correlations are preserved. We take the median as orange curves shown in Fig.~\ref{fig:Gs} and \ref{fig:ForceIMF}, and uncertainties are estimated using $1\sigma$ confidence intervals, shown as  orange shaded areas. 

On the other hand, a different parametrization is adopted for the EMT FFs extracted from the experimental data. The $A_q(q^2)$-form factor is obtained from the global extraction of generalized parton distributions, or the GUMP fit~\cite{Guo:2025muf}, while the $C_q(q^2)$-form factor is extracted from BEG's dispersive analysis of DVCS data~\cite{Burkert:2018bqq}.

In the GUMP fit, the conformal moments of generalized parton distributions are parametrized, where the first conformal moment of $H_q(x,\xi,q^2)$ reduces to the familiar Mellin moment, and thus EMT FFs~\cite{Guo:2025muf}, 
\begin{align}
    \int_{-1}^1 {\rm d}x~xH_q(x,\xi,q^2) &= A_q(q^2) + 4\xi^2C_q(q^2) \equiv {\cal F}_{1,0}^q (q^2)+ \xi^2 {\cal F}_{1,2}^q (q^2) \ ,
\end{align}
where ${\cal F}_{1,0}^q(q^2),{\cal F}_{1,2}^q(q^2)$ are parametrized and correspond to $A_q(q^2),C_q(q^2)$ form factors, respectively. Note that since the $\xi$ dependence in the above moment is neglected---${\cal F}_{1,2}^q$ is not implemented in the parameterization---it only yields the $A_q$-form factor, which is parametrized as,
\begin{equation}
    A_q^f(q^2)={\cal F}_{1,0}^f(q^2) = \sum_{i=1}^{i_{\max}}N_{i,f} \frac{B(2-\alpha_{i,f}-\alpha_{i,f}^{\prime}q^2,1+\beta_{i,f})}{B(2-\alpha_{i,f},1+\beta_{i,f})}R_{i,f}(q^2)
\end{equation}
where $f=u,d,\bar{u},\bar{d},g$ represents the flavor, $B(x,y)$ is the beta function, $\{N_{i,f},\alpha_{i,f},\beta_{i,f},\alpha_{i,f}^{\prime}\}$ are parameters obtained by  the GUMP fit, $i_{\max}=1$ for valence quarks $u,d$ and $i_{\max}=2$ for sea quarks and gluons $\bar{u},\bar{d},g$, $R_{i,f}(q^2)=(1-q^2/M_{i,f}^2)^{-2}$ for $u,d$ and $R_{i,f}(q^2)=\exp(-b_{i,f}q^2)$ for $\bar{u},\bar{d},g$. The total $A_q(q^2)$-form factor is obtained by summing over all of the  quark flavors. 

The $C_q$-form factor is extracted by the BEG fit using a dispersive analysis of the DVCS data of beam-spin asymmetry and unpolarized cross sections. This form factor is also parametrized using a multipole ansatz, except that the power $\alpha$ is also a parameter to be fitted. 

Combining the $A_q(q^2)$-form factor from the GUMP fit and $C_q(q^2)$-form factor from the BEG fit as well as the Dirac form factors, we plot the quark scalar form factor in Fig.~\ref{fig:Gs} and the quark force in Fig.~\ref{fig:ForceIMF} as the black curves. Uncertainties are estimated by adding errors in parameters in quadrature. Specifically, for a function $f(r;\{A_i\})$ depending on several parameters $\{A_i\}$ with corresponding errors $\{\Delta A_i\}$, the $1\sigma$ uncertainties are,
\begin{equation}
    \Delta f(r;\{A_i\}) = \sqrt{\sum_i\left(\frac{\partial f(r;\{A_i\})}{\partial A_i}\Delta A_i\right)^2}
\end{equation}
In the case of quark scalar form factor and force density, since the error analysis is not implemented in the current GUMP fit, uncertainties arise mainly from the $C_q(q^2)$-form factor extracted from BEG's fit, shown as the gray shaded area in Fig.~\ref{fig:Gs} and \ref{fig:ForceIMF}. To account for possible underestimated errors in BEG's fit, we add additional 20\% errors in these parameters.

\end{document}